\documentclass[10pt,conference]{IEEEtran}
\usepackage{amsmath,amssymb,mathrsfs}
\usepackage{epsfig,epsf,subfigure,graphicx,graphics}
\usepackage{url}
\usepackage{color}

\newtheorem{theorem}{Theorem}[section]

\graphicspath{{../figs/}}

\IEEEoverridecommandlockouts

\allowdisplaybreaks

\usepackage{cite,subfigure}
\usepackage{algorithmic,algorithm}
\usepackage{xcolor}
\usepackage{epstopdf}

\title{Embedding Information \\in Radiation Pattern Fluctuations}
\author{
\authorblockN{Milad Johnny} \and
\authorblockN{Alireza Vahid} 
\thanks{The extended version of this work is submitted to IEEE Transactions on Communications~\cite{FluctuationsJournal}. Alireza Vahid is with the Department of Electrical Engineering at the University of Colorado Denver, Denver, USA. Authors’ email
addresses: {\sffamily mjohnny@uwaterloo.ca, \sffamily alireza.vahid@ucdenver.edu}. } 
}

\begin{document}
\maketitle


\begin{abstract}
The radiation pattern of transmit antennas varies and fluctuates as receivers change their location, other objects move around, and due to the antenna design itself. In this paper, we demonstrate how this observation can be exploited to align most of the interference signal power and significantly increase the average achievable communication rates. More precisely, in the context of $K$-user interference channels, we propose a blind interference alignment scheme that combines multi-layer coding at the transmitters and a post-processing methodology at the receivers to align a significant portion of the interference signal power. Our scheme does not rely on any channel state information (CSI), hence the term blind, and only relies on the statistics of the radiation pattern fluctuations. Our proposed communication methodology overcomes some of the barriers in practical implementation of the interference alignment concept. Due to the complexity of the expressions, in this work, we numerically evaluate the achievable rates in different scenarios, demonstrate the gains of our proposed strategy, and compare our results to the prior works with perfect CSI.  
\end{abstract}



\section{Introduction}
\label{Section:Introduction_BIA}

In recent years, alleviating the negative impact of interference in wireless systems has been at the center of attention. From a physical-layer point of view, there are well-known interference avoidance methods such as TDMA and FDMA that reduce the impact of interference at the receivers at the cost of spectrum efficiency. Considering the ever-growing demand for higher communication rates and the exponential growth of wireless users, we need to replace these suboptimal methods with more bandwidth-efficient schemes. Interference alignment (IA) is an attractive concept in wireless networks which creates a multiplexing gain proportional to the number of users in the network~\cite{CadambeJafar}. However, the requirement of perfect CSI, and in some cases global perfect CSI, the fast fading assumption, and long precoder lengths at the transmitters are some of the main barriers in implementing IA~\cite{choi2009beamforming,el2013practical,johnny2017efficient}. 

Broadly speaking, in terms of reliance on CSI, we can categorize IA schemes into four groups: The first one is based on perfect channel state information (CSI)~\cite{alignmentKhandani,CadambeJafar,peters2009interference,suh2011downlink,gollakota2009interference}; the second one is instantaneous but imperfect CSI~\cite{nosrat2011mimo,CadambeJafar,el2011interference}; the third one is delayed CSI~\cite{MAT,retrospective,AlirezaBFICDelayed,Castanheira,vahid2019degrees,vaze2012degrees,vahid2016approximate,vahid2016two}; the last one is known as blind IA (BIA)~\cite{jafar2012blind,BIAJohnny2018,BIAJohnny2017,Johnny2019blind,gou2011aiming,FluctuationsJournal}. For the first three cases of (perfect, imperfect, and delayed) CSI, one of the main practical challenges is attaining high-resolution CSI cannot be realized (with acceptable overhead) in real-world networks~\cite{el2013practical}. The last one, the BIA, does not rely on CSI at the transmitters but typically assumes particular channel variation structures that may not be feasible in practice.

In this paper, we propose a new take on IA technique based on the observation that the radiation pattern of transmit antennas varies and fluctuates as receivers change their location, other objects move around, and due to the antenna design itself. In fact, in~\cite{FluctuationsJournal}, we propose a transmit antenna hardware design that creates such an environment even when wireless nodes are stationary. Thus, our main contribution in this paper is the introduction of a new BIA scheme based on antenna radiation pattern fluctuations which result in different coherence times for different wireless links. Using the statistics of channel coherence times, we introduce a multi-layer encoding, a post-processing, and a successive decoding strategy to maximize the average transmission rate. We show that without accessing CSI and even channel variations pattern, we can improve the average achievable sum-rate drastically similar to prior IA schemes (even those with perfect CSI). 

In the next section, we introduce our system and channel models, and explain how different receivers may experience different channel coherence times from different transmitters. In Section~\ref{Section:Main_IFB}, using differences in channel coherence times, without accessing channel state information at the transmitters, we devise our BIA scheme. Finally, through numerical analysis, we show our scheme increases the average transmission rate, and we compare our results to known outer-bound.


\section{Problem Formulation}
\label{Section:Problem_BIA}

In this paper, we consider the $K$-user interference channel (IC) in which $K$ single-antenna transmitters $\mathrm{TX}_k, k \in \mathcal{K}=\{1,\dots,K\}$, send their messages to $K$ single-antenna receivers $\mathrm{RX}_k, k \in \mathcal{K}$. The received signal at $\mathrm{RX}_i$ is given by
\begin{equation}
{\bar{\bf Y}}^{[i]}= \sum_{j=1}^{K}{{\bar {\bf H}}^{[ij]} {\bf {\bar X}}^{[j]}}+\bar{\bf Z}^{[i]}, \qquad i, j \in \mathcal{K},
\end{equation}
where the $n \times 1$ column vectors ${\bar{\bf Y}}^{[i]}$ and ${\bf {\bar X}}^{[j]}$ represent the received and the transmitted signals of $\mathrm{RX}_i$ and $\mathrm{TX}_j$, respectively. Here, $n$ is the communication block length. The $n \times 1$ column vector $\bar{\bf Z}^{[i]}$ denotes the additive white Gaussian noise at $\mathrm{RX}_i$ distributed according to $\mathcal{CN}\left(0,1\right)$, and $\bar{\bf H}^{[ij]}=\mathrm{diag}\left({\left[{h^{[ij]}_{1},\dots,h^{[ij]}_n}\right]}\right), i,j \in \mathcal{K}$, is the  $n\times n$ diagonal matrix of channel coefficients from $\mathrm{TX}_j$ to $\mathrm{RX}_i$.
As motivated later in this section, and due to transmitter antenna radiation pattern fluctuations, we assume a block fading model in time, where channels remain constant for some duration, and these durations may vary for different channels. Therefore, for the channel matrix of $\bar{\bf H}^{[ij]}$, we have:
\begin{equation}
\label{channel1}
{h}_{{c_{l}^{[ij]}}}^{[ij]}= \ldots ={h}_{{c_{l+1}^{[ij]}-1}}^{[ij]}
\end{equation}
where $l \in \{1,\dots,\eta \left({i,j}\right)\}$ and $\eta \left({i,j}\right)$ is the number of channel ``altering points'' in time between $\mathrm{TX}_j$ and $\mathrm{RX}_i$. The value of $c_{l}^{[ij]},~l\in \{1,\dots,\eta \left( i,j \right)\}$ represents the $l^{th}$ point of altering state of the channel between $\mathrm{TX}_j$ and $\mathrm{RX}_{i}$ and as $n$ goes to infinity $\frac{\eta \left( i,j \right)}{n}$ represents channel variation rate. It is assumed that ${h}_{{c_{l}^{[ij]}}}^{[ij]}$'s are $i.i.d$ random variables with a specific distribution with {\it magnitude} bounded between a nonzero value and a finite maximum value. Since the channel coherence time is a random variable, the channel at each altering point in time may remain in its previous state with probability $p_{ij}$, and may transition to a new state with probability $(1-p_{ij})$. For the channel described above, we have the following definition.

{\it{Definition~1:} We define $\mathcal{F}\left({\bar{\bf H}}\right)$ to be the maximum number of the time snapshots in which channel matrix ${\bar{\bf H}}$ remains constant. We refer the reader to~\cite{FluctuationsJournal} for further intuitions on this definition.}


In this paper, we use a multi-layer encoding and decoding strategy in which transmitter $\mathrm{TX}_i$ wishes to send uniformly distributed message $W^{[i]} \in \mathcal{W}^{[i]}=\mathcal{W}^{[i]}_1 \times \dots \times \mathcal{W}^{[i]}_M$ to $\mathrm{RX}_i$, $i \in \mathcal{K}$, over $n$ uses of the channel. Each set $\mathcal{W}^{[i]}_j$, $1 \leq j \leq M$, represents the message set for the $j^{\mathrm{th}}$ transmission layer. We further assume that the messages are independent from each other and the channel gains. Each transmitter is subject to a total average transmission power constraint of $P_t$. Transmitter $\mathrm{TX}_i$ encodes its message $W^{[i]}$ using the encoding function ${\bf {\bar X}}^{[i]}= e_i \left({W^{[i]}}, \mathrm{SI}_{{\sf TX}_i} \right)$ where $i \in \mathcal{K}$ and $\mathrm{SI}_{{\sf TX}_i}$ is the available side information at $\mathrm{TX}_i$ which is a cumulative distribution function discussed in Section~\ref{Section:Decoding}. The value of $R^{[i]}_j={\log_{2}{\lvert {\mathcal{W}^{[i]}_j}\rvert}}/{n}$ is the transmission rate of of $j^{\mathrm{th}}$ layer at $\mathrm{TX}_i$. Therefore, the total transmission rate at $\mathrm{TX}_i$ is $R^{[i]}=\sum^{M}_{j=1}{R^{[i]}_j}$. We assume that each receiver $\mathrm{RX}_i$ is aware of its channel state information and decodes its intended message $W^{[i]}_j \in \mathcal{W}^{[i]}_j$ using the decoding function $\hat{W}^{[i]}_j= \phi_{ij} \left( {\bar{\bf Y}}^{[i]}, \mathrm{SI}_{{\sf Rx}_i} \right)$, $1\leq j \leq M$, where $\mathrm{SI}_{{\sf Rx}_i}$ is the side information available to the receiver (in this case channel state information). Then, the decoding error probability at receiver $\mathrm{RX}_i$ for the $j^{\mathrm{th}}$ layer is given by
\begin{align}
\lambda^{[i]}_{j}(n) = \mathbb{E}\left[\Pr \left( \hat{W}^{[i]}_j \neq {W}^{[i]}_j \right)\right],
\end{align}
where the expectation is over the random choice of messages. For $\mathcal{J}_i \subseteq \{1,\dots,M\}$, we define 
$R^{[i]}_{\mathcal{J}_i} = \sum_{j \in \mathcal{J}_i}{R^{[i]}_j}$.
Then, rate-tuple $\left( R^{[1]}_{\mathcal{J}_1}, R^{[2]}_{\mathcal{J}_2}, \ldots, R^{[K]}_{\mathcal{J}_K} \right)$, is achievable if there exist encoders and decoders such that $\lambda^{[i]}_{j}(n) \rightarrow 0$,  when $n \rightarrow \infty$ for all $i \in \mathcal{K}$, and $j \in \mathcal{J}_i$.  Based on this multi-layer definition for encoding and decoding, we are able to present and evaluate the average achievable rates in Section~\ref{Section:AchRates}.

\begin{figure}
  \centering
  \includegraphics[width=\columnwidth]%
    {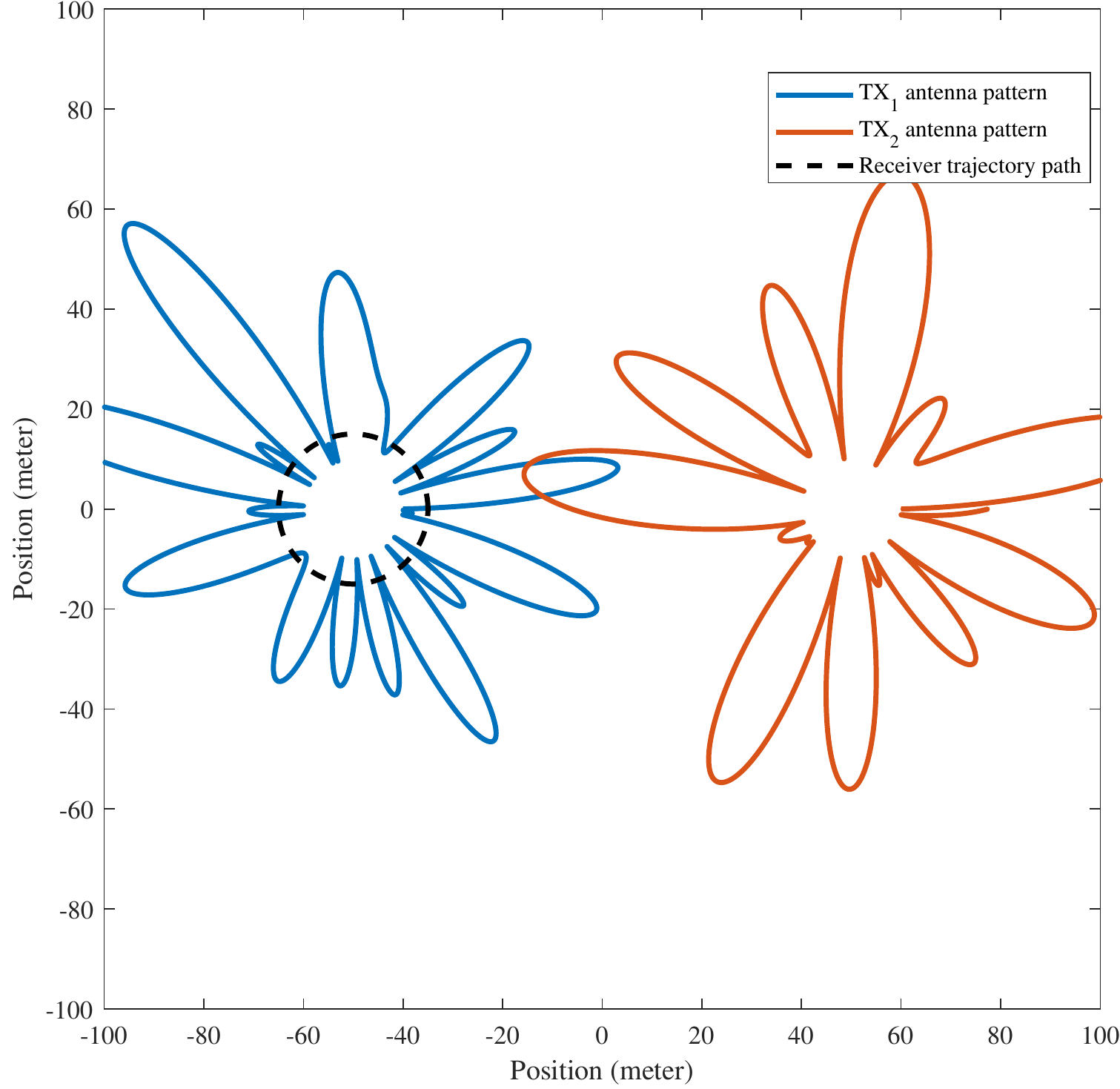}
  \caption{A receiver moves around its corresponding transmitter and a second nearby transmitter creates interference at this receiver. The receiver experiences higher channel gain fluctuations from the corresponding transmitter.}
  \label{RFregion}
\end{figure}

We further limit our study to the scenario in which cross channels vary at a much slower pace compared to direct channels, and a more general setting is considered in the extended version of this work~\cite{FluctuationsJournal}. Our channel model is motivated by the physics of wireless networks. Consider an interference channel in which receivers have random and time-varying physical locations, and each transmitter is equipped with an antenna with high fluctuation pattern gain. In Fig.~\ref{RFregion}, a receiver is in the vicinity of two transmitters (the intended and the interfering transmitters). The intended transmitter is typically closer to its receiver, and thus, the receiver observes a higher fluctuation rate for the intended signal. 
 
\section{Opportunistic Blind Interference Alignment}
\label{Section:Main_IFB}

We now present our transmission strategy that opportunistically utilizes the channel variation pattern in order to align a part of interference signal power at the receivers. We note that the transmitters do not have access to the channel state information or the channel variation knowledge, and are only aware of the statistics of the channel parameters. We will deploy multi-layer superposition coding at the transmitters  to maximize the average achievable sum-rate. More precisely, transmitters $\mathrm{TX}_{k}, 1\leq k \leq K$, use a multi-layer encoding scheme with $M$ distinct layers $l^{[k]}_i , 1 \leq i \leq M$. In this coding procedure, the transmit signal in each layer $l^{[k]}_i , 1 \leq i \leq M$ is represented by a Gaussian random variable with the length of $n$. As discussed later, these transmission layers are designed such that at each receiver, the decoder (based on the fluctuation pattern) is able to recover some parts of the transmitted data.


To present our Interference Alignment technique, we need the following lemma whose proof is presented in~\cite{FluctuationsJournal}.

{\it{Lemma 1}: Consider $\bf V$, a random matrix with nonzero elements of size $n \times d_v$ and rank $d_v$. Let $\bar{\bf H}$ be a random diagonal matrix with size $n \times n$, we have:
\begin{align*}
\mathrm{rank}\left({\left[{{\bf V~\bar{\bf H}{\bf V}}}\right]}\right) \overset{a.s.}= \min\left({{d_v+\min{\left({d_v,n-\mathcal{F}\left({\bar{\bf H}}\right)}\right)}},n}\right),
\end{align*}
where $\left[{{\bf V~\bar{\bf H}{\bf V}}}\right]$ is obtained by concatenating ${\bf V}$ and $\bar{\bf H}{\bf V}$.}

\subsection{Encoding}
 
The encoding strategy has the following steps:\\
\noindent $(1)$ Let $X^{[k]}_{i}$, $1 \leq i \leq M$, be $M$ Gaussian distributed continuous random variables with limited variance for the $i^{th}$ transmission layer at $\mathrm{TX}_k$.

\noindent $(2)$ At $\mathrm{TX}_k$, we partition our message $W^{[k]}$ into $M$ sub-messages of $\left({W^{[k]}_1,\cdots,W^{[k]}_M}\right)$, and we send each sub-message via a single transmission layer. 
For each layer $l^{[k]}_i, 1 \leq i \leq M$, consisting of $\frac{n}{2}$ time snapshots, we consider a $\left({2^{n R^{[k]}_i},\frac{n}{2}}\right)$ code for the Gaussian channel where $W^{[k]}_i \in \{1,\dots,2^{n R^{[k]}_i}\}$, and we generate codeword of $x^{[k]\frac{n}{2}}_{i1} (W^{[k]}_i)$ as follows:
\begin{equation}
x^{[k]\frac{n}{2}}_{i1} \left({W^{[k]}_i}\right)=\left({x^{[k]}_{i1}({W^{[k]}_i}),\dots,x^{[k]}_{i \frac{n}{2}}\left({W^{[k]}_i}\right)}\right)^{\intercal}.
\end{equation}

\noindent $(3)$ $\mathrm{TX}_k$ computes $X^{[k]}_j=\sum_{i=1}^{M}{x^{[k]}_{ij} (W^{[k]}_i)}$ of $M$ independent layers to generate the following matrix:
\begin{equation}
 {\bar{\bf{x}}}^{[k]}=\left[{X^{[k]}_1, \dots, X^{[k]}_{\frac{n}{2}}}\right]^{\intercal}.
\end{equation}

\noindent $(4)$ Transmitters use a random full rank matrix of $\bar{\bf{V}}$ with the size of $n \times \frac{n}{2}$ as their precoder matrices. We can design $\bar{\bf{V}}$ such that $\lim_{n \to \infty}{\frac{1}{n}\mathrm{tr}\left({\bar{\bf{V}} {\bar{\bf{x}}}^{[k]}(\bar{\bf{V}} {\bar{\bf{x}}}^{[k]})^{H}}\right)}=P_t$, where $\mathrm{tr}\left({\bf A}\right)$ computes the trace of the square matrix $\bf A$.

\noindent $(5)$ The output of the encoding function at $\mathrm{TX}_k$, ${{\bf \bar{X}}^{[k]}}$, is an $n \times 1$ column vector with the following relation:
\begin{equation}
\label{3}
{{\bf \bar{X}}^{[k]}} =\bar{\bf{V}} {\bar{\bf{x}}}^{[k]}.
\end{equation}
 and for $i^{th}$ layer we have the following constraint:
\begin{equation}
 \lim_{n \to \infty}{\frac{1}{n}\mathrm{tr}\left({\bar{\bf{V}} x^{[k]\frac{n}{2}}_{i1}\left({W^{[k]}_i}\right) \left({\bar{\bf{V}} x^{[k]\frac{n}{2}}_{i1}\left({W^{[k]}_i}\right)}\right)^{H}}\right)}= P^{[k]}_i,  
\end{equation}
where $P^{[k]}_i$ indicates the total transmission power for $i^{\mathrm{th}}$ layer at $\mathrm{TX}_k$ and $\sum_{i=1}^{M}{P^{[k]}_i}=P_t$, which shows that the summation of transmission power for all of the layers is equal to $P_t$.

\subsection{Analyzing the received signal}

\begin{figure}
\label{receptionspace1}
  \centering
  \includegraphics[width=\linewidth]%
    {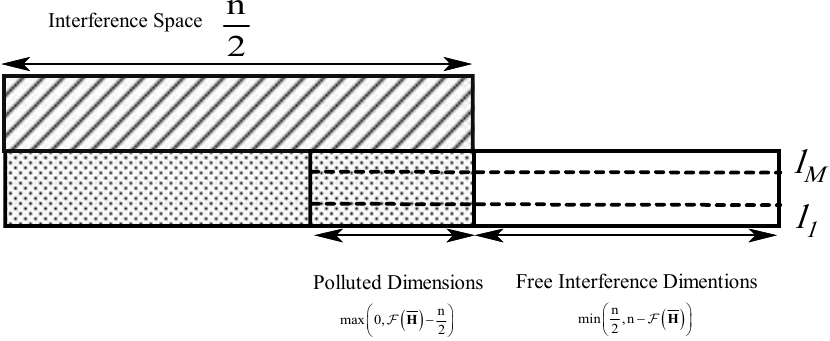}
  \caption{The reception space at each receiver, prior to applying the zero-forcing decoding matrix, consists of three different subspaces: the desired interference-free subspace of dimension $\min {\left({\frac{n}{2}, n- \mathcal{F}{{(\bar{\bf H})}}}\right)}$; the desired signal sub-space corrupted by interference signals of dimension  $\max{\left({0,  \mathcal{F}{{(\bar{\bf H})}}-\frac{n}{2}}\right)}$; the remaining interference sub-space. Note that we assume that all cross links have constant values.}
  \label{receptionspace}
\end{figure}

Fig.~\ref{receptionspace} shows a representation of the reception space at a receiver in a 3-user IC. In this case, we assume $p_{ij}>>p_{ii}, i \neq j$, meaning that the probability of changing channel value for the cross channels is much lower than the direct channels. The goal is to align the interference signals at each receiver, and depending on the value of $\mathcal{F}\left({\bar {\bf H}}\right)$, the direct channel signal space has some free interference dimensions which are linearly independent from the interference signals. The decoder uses the following zero-forcing decoding matrix:
\begin{equation}
{\bar{\bf D}}=\left({{\bar{\bf I}}-{\bar{\bf V}}({\bar{\bf V}}^{H}{\bar{\bf V}})^{-1}{\bar{\bf V}}^{H}}\right).
\end{equation}
If we assume during $n$ transmission time snapshots, all cross links have a constant diagonal value, we have $\bar{\bf D}\bar{\bf H}^{[ij]} {\bf{V}}{\bar{\bf{x}}}^{[j]}=\bar{\bf H}^{[ij]} \bar{\bf D} {\bf{V}}{\bar{\bf{x}}}^{[j]}={\bf{0}}$.
Since all transmitters use the same precoder of $\bar{\bf{V}}$, for the encoding scheme of the previous subsection where all transmitted signals have the same number of dimensions $\frac{n}{2}$, the number of free interference subspace dimensions at each receiver can be calculated from Lemma~1 as:
\begin{align}
&\mathrm{rank}\left({\left[{{\bf V}~\bar{{\bf H}}{\bf V}}\right]}\right)-\mathrm{rank}\left({\left[ {\bf V} \right]}\right)&\nonumber\\
&\overset{a.s.}=\min\left({{d_v+\min{\left({d_v,n-\mathcal{F}\left({\bar{\bf H}}\right)}\right)}},n}\right)-d_v,&
\end{align}
substituting $d_v=\frac{n}{2}$ in the above equation, we have:
\begin{align}
&\mathrm{rank}\left({\left[{{\bf V}~{\bar{\bf H}}{\bf V}}\right]}\right)-\mathrm{rank}\left({\left[ {\bf V} \right]}\right)\nonumber\\
&\overset{a.s.}=\min\left({{\frac{n}{2}+\min{\left({\frac{n}{2},n-\mathcal{F}\left({\bar{\bf H}}\right)}\right)}},n}\right)-\frac{n}{2}\\
&=\min\left({{\min{\left({\frac{n}{2},n-\mathcal{F}\left({\bar{\bf H}}\right)}\right)}},\frac{n}{2}}\right)&\\
\label{Eq:MinFreeSpace}
&=\min{\left({\frac{n}{2}, n-\mathcal{F}\left({\bar{\bf H}}\right)}\right)},&
\end{align}
where in the above relations $\bar{\bf{H}}$ represents the direct channel matrix.

\subsection{Decoding}
\label{Section:Decoding}

Without loss of generality, we consider the decoding operation at the first receiver. As mentioned earlier, we assume all channel matrices are known to their corresponding receivers. Furthermore, for simplicity and in order to provide a comparison to prior results that focus on DoF,  we assume $\mathrm{det}\left({\bar{\bf H}\bar{\bf H}^{H}}\right)=1$. In other words, in this work, we ignore the potential gain of the exact knowledge of channel gain values and assume channel matrices give no power gain to the received signals during transmission block.
With slight abuse of notation, we denote the set of jointly $\epsilon$-typical sequences of length $n$ by $\mathcal{T}^{n}_{\epsilon}$. 

After applying zero forcing matrix, $\bar{\bf{D}} {\bf Y}^{[1] n}_1$ is available at the first receiver. Based on the following successive decoding strategy, we can decode part of data:  

\noindent $(1)$ First layer decoding: Based on $\bar{\bf H}^{[ij]}$ and the decoder matrix of $\bar{\bf D}$, the decoder finds that $\widehat{w}_1$ is sent if there exists a unique message such that 2-tuple $n-$sequence of $\left({{\bar{\bf D}}{\bar{\bf H}^{[11]}}{\bar{\bf V}}x^{[1]\frac{n}{2}}_{11}{\left({\widehat{w}_1}\right)},\bar{\bf{D}} {\bf Y}^{[1] n}_1}\right) \in \mathcal{T}^{n}_{\epsilon}$, otherwise it declares error in the first decoding step. 

\noindent $(2)$ Second layer decoding: After decoding the first layer and if such $\widehat{w}_1$ is found, the decoder tries to find $\widehat{w}_2$ such that the 3-tuple $n-$sequence of $\left({{\bar{\bf D}}{\bar{\bf H}^{[11]}}{\bar{\bf V}}x^{[1]\frac{n}{2}}_{11}{\left({\widehat{w}_1}\right)},{\bar{\bf D}}{\bar{\bf H}^{[11]}}{\bar{\bf V}}x^{[1]\frac{n}{2}}_{21}{\left({\widehat{w}_2}\right)},\bar{\bf{D}} {\bf Y}^{[1] n}_1}\right) \in \mathcal{T}^{n}_{\epsilon}$, otherwise it declares error in the second decoding stage.
 
\noindent $(3)$ $i$-th layer decoding: After finding unique messages ${\widehat{w}_1},\cdots,{\widehat{w}_{i-1}}$ in the previous steps, the decoder finds $\widehat{w}_i$ such that $(i+1)-$tuple of $n-$sequences $\left({{\bar{\bf D}}{\bar{\bf H}^{[11]}}{\bar{\bf V}} x^{[1]\frac{n}{2}}_{11}{\left({\widehat{w}_1}\right)},\ldots,{\bar{\bf D}}{\bar{\bf H}^{[11]}}{\bar{\bf V}}x^{[1]\frac{n}{2}}_{i1}{\left({\widehat{w}_i}\right)},\bar{\bf{D}} {\bf Y}^{[1] n}_1}\right) \in \mathcal{T}^{n}_{\epsilon}$, otherwise it declares error in the $i$-th decoding step. 

The error analysis of this decoding approach is presented in~\cite{FluctuationsJournal} where we show we can set the following transmission rate for $i$-th transmission layer:
\begin{equation}
\label{eq:rate_relation}
n R^{[1]}_i \leq \frac{i}{2} \log{\left({1+\frac{P^{[1]}_i}{N+\sum_{j=i+1}^{M}{P^{[1]}_j}}}\right)},~M=\frac{n}{2}
\end{equation} 
where $N$ is the noise power, and we treat interference from all layers that are not yet decoded as noise, also the value of $i$ indicates the number of free interference dimension at the receiver which is appeared in multiplication form in the above relation. Each layer which is decoded at receiver should be subtract from the received signal to decode the next layer.\\
{\it{Note:} We assume that the channel matrices and the zero-forcing operation add no gain to decoding or have the same impact on signals and noise. Therefore, for the sake of simplicity, we omit their impact in our analysis.}

Depending on channel realizations, each receiver will be able to decode part of its intended data. More precisely, if the number of interference-free dimensions at the receiver is $l_s$, we except the receiver to be able to decode the messages of layers $1$ to $l_s$. Based on~\eqref{Eq:MinFreeSpace}, in the next subsection, we will divide the total transmission power among $M=n/2$ layers in order to maximize the average transmission rate.  

\subsection{Maximizing the average transmission rate}
\label{Section:AchRates}

Based on the argument presented in the previous subsection, the average transmission rate is given by:
\begin{equation}
\label{Arate}
\bar{R}_{\mathrm{avg}}=\frac{1}{n}\sum_{i=1}^{\frac{n}{2}}{\frac{i}{2} \log{\left({1+\frac{P^{[1]}_i}{N+\sum_{j=i+1}^{M}{P^{[1]}_j}}}\right)}F_{I}\left({\frac{n}{2}-i}\right)}
\end{equation} 
where 
\begin{align}
\label{Eq:TXSideInformation}
F_{I}\left({\frac{n}{2}-i}\right)=p\left(\min{\left({\frac{n}{2}, n-\mathcal{F}\left({\bar{\bf H}}\right)}\right)} \geq i\right). 
\end{align}
For large $n$, $\frac{i}{n}, 1 \leq i \leq \frac{n}{2}$ and $P^{[1]}_i$ can be approximated by continues parameters of $0 \leq z \leq \frac{1}{2}$ and $\rho (z) dz$, respectively.
 
We set $P(1/2-z)=\int_{z}^{\frac{1}{2}}{\rho (x) dx}$, and thus, $P\left({0}\right)=\int_{\frac{1}{2}}^{\frac{1}{2}}{\rho (x) dx}=0$ and $P\left({1/2}\right)=\int_{0}^{\frac{1}{2}}{\rho (x) dx}=P_t$. Therefore, for the large values of $n$, \eqref{Arate} can be approximated as:
 \begin{equation}
 \label{Arate2}
\bar{R}_{\mathrm{avg}}=\frac{1}{2 \ln{2}} \int_{0}^{\frac{1}{2}}{\frac{z \rho (1/2-z) F_Z \left({1/2-z}\right)}{N+P(1/2-z)}dz}
\end{equation}
where $F_Z \left({{1}/{2}-z}\right)=F_{I}\left({n}/{2}-{\lfloor{n z}\rfloor}\right),~0 \leq z \leq {1}/{2}$ and $\rho (z) ={d P(z)}/{dz}$. We note that as the number of layers $M=n/2$ increases, our approximation becomes more accurate.

The following theorem helps us maximize the above average transmission rate. For simplifying our notation, we substitute $\frac{1}{2}-z$ with $u$ to get:
\begin{align}
 \label{Arate2}
\bar{R}_{\mathrm{avg}}&=\frac{1}{2 \ln{2}} \int_{\frac{1}{2}}^{0}{\frac{-(1/2-u) \rho (u) F_U \left({u}\right)}{N+P(u)}du}&\\
&=\frac{1}{2 \ln{2}} \int_{0}^{\frac{1}{2}}{\frac{(1/2-u) \rho (u) F_U \left({u}\right)}{N+P(u)}du}.&
\end{align}

\begin{theorem}
\label{thm_1} \it{A necessary condition for function $y(u)$ to be an extremum of:
\begin{equation}
\int_{z_1}^{z_2}{D\left({y,y^{'},u}\right)du},
\end{equation}
is that $y$ satisfies the following Euler differential equation:
\begin{equation}
D_{y}-\frac{d D_{y^{'}}}{du}=0,~{z_1}\leq u \leq {z_2}.
\end{equation} 
where the subscripts denote the partial derivatives with respect to corresponding arguments~\cite{Gelfand,johnny2019multi}.}
\end{theorem}

\begin{figure}
  \centering
  \includegraphics[width=\columnwidth]%
    {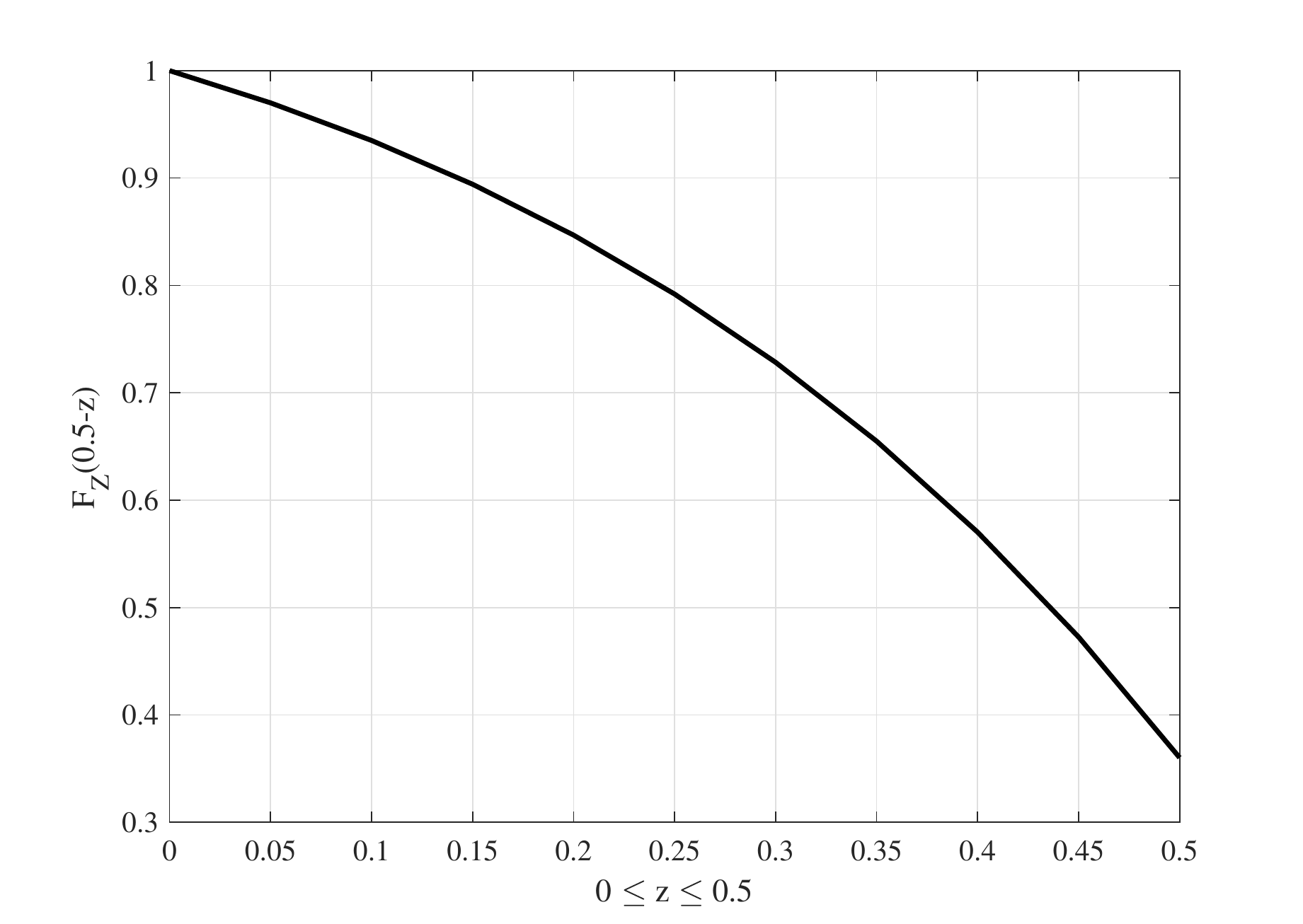}
  \caption{Numerical evaluation of the cumulative distribution function $F_{Z}(z)$ from $F_{I}(i)$. See~\cite{FluctuationsJournal} for more details.}
  \label{cummulativedistributiong}
\end{figure}

\begin{figure}
  \centering
  \includegraphics[width=\columnwidth]%
   {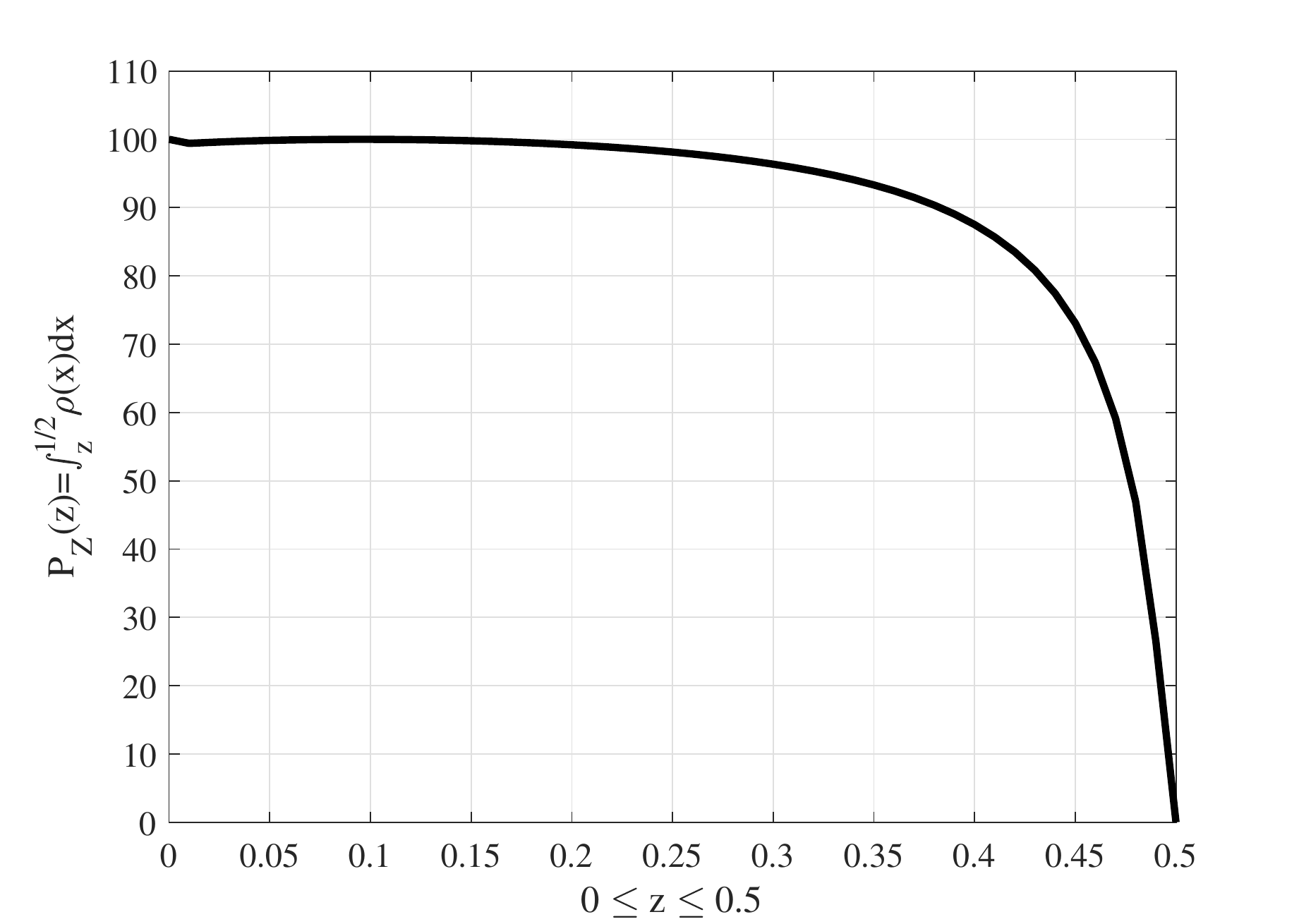}
  \caption{$P_{Z}(z)$ as a function of $0 \leq z \leq \frac{1}{2}$ using $F_{Z}(z)$ in Fig.~\ref{cummulativedistributiong} for in which $p_{11}=0.9$ and $P^{[1]}_t=100$.}
  \label{commulativepower}
\end{figure}

\begin{figure}[ht]
  \centering
  \includegraphics[width=\columnwidth]%
    {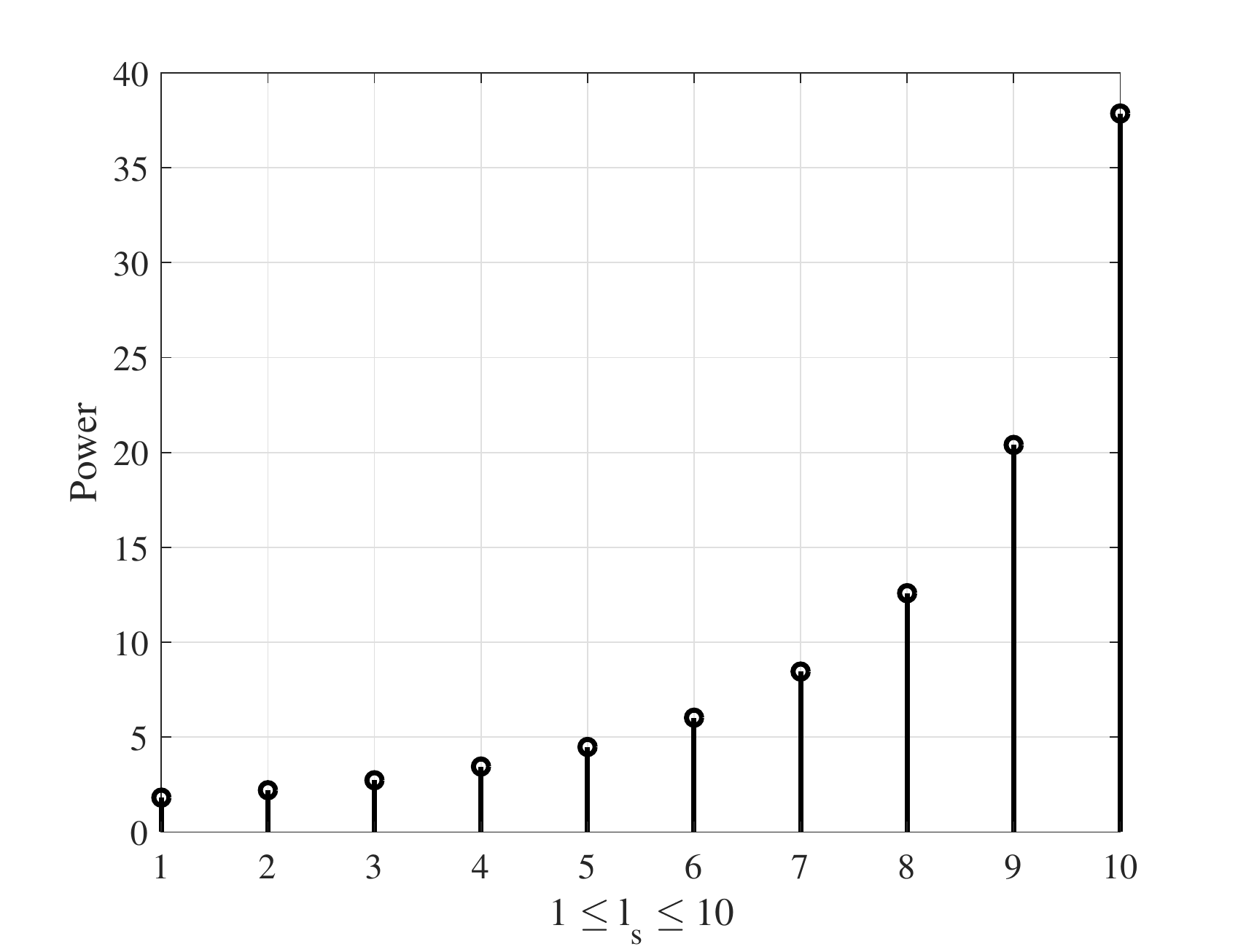}
  \caption{Dedicated power to each transmission layer for $n=20$ and $p_{11}=0.9$.}
  \label{Powerforeachlayer}
\end{figure}

In our problem, we have:
\begin{equation}
D(y,y',u)=\frac{1}{2 \ln {2}}\frac{(1/2-u) \rho (u) F_U \left({u}\right)}{N+P(u)}
\end{equation}
 where $y(u)=P(u)$ and $y'(u)=\rho(u)$. We have:
 \begin{align}
 &D_y =\frac{1}{2 \ln 2}\frac{-(1/2-u) \rho (u) F_{U}(u)}{\left({N+P(u)}\right)^2}&\\
 &D_{y'}=\frac{1}{2 \ln 2}\frac{(1/2-u)  F_{U}(u)}{{N+P(u)}}.&
 \end{align}
Thus, we get 
 \begin{equation}
 \frac{d D_{y^{'}}}{du}=\frac{(-F_{U}(u)+(1/2-u) f_{U}(u))(N+P(u))}{2 \ln 2 (N+P(u))^2},
 \end{equation}
where $f_{U}(u)=\frac{d F_{U}(u)}{du}$, and from the first theorem we have:
 \begin{equation}
 \frac{d P(u)}{N+P(u)}=-\frac{(-F_{U}(u)+(1/2-u)f_{U}(u)) du}{(1/2-u)F_{U}(u)},
 \end{equation}
 and finally:
 \begin{equation}
 \ln {\left({N+P(u)}\right)}=-\ln {\left({(1/2-u) F_{U}(u)}\right)}+C.
 \end{equation}
 Therefore:
 \begin{equation}
 P(1/2-z)=\frac{C}{ z F_{Z}(1/2-z)}-N,
 \end{equation}
 where $P(1/2-z=0)=0$. Setting $z=1$, we get $C={N~F_{Z}\left({0}\right)}/2$,
 and function $P(z)$ can be calculated as:
 \begin{equation}
 P(z)=\frac{N}{2 (1/2-z) F_{Z}(z)} F_{Z}\left({0}\right)-N.
 \end{equation}
We note that when $P(z_0)\geq P_t$ then $P(z>z_0)=P_t$. Fig. \ref{cummulativedistributiong} depicts numerical evaluation of $F_{Z}(z)$ for precoder length of $n=20$, and the altering probability of $p_{ii}=0.9$. Given $F_{Z}(z)$, we can calculate $P(z)$ (as in Fig. \ref{commulativepower}), and then, the transmission power for each transmission layer can be calculated as:
 \begin{equation}
 P^{[1]}_j=P_Z\left({\frac{j-1}{n}}\right)-P_Z\left({\frac{j}{n}}\right).
 \end{equation}
 For $P_t=100$, $n=20$ and $N=1$, the total transmission power for each transmission layer is depicted in Fig.~\ref{Powerforeachlayer}. 

\subsection{Numerical Results}

\begin{figure}[ht]
  \centering
  \includegraphics[width=\columnwidth]%
    {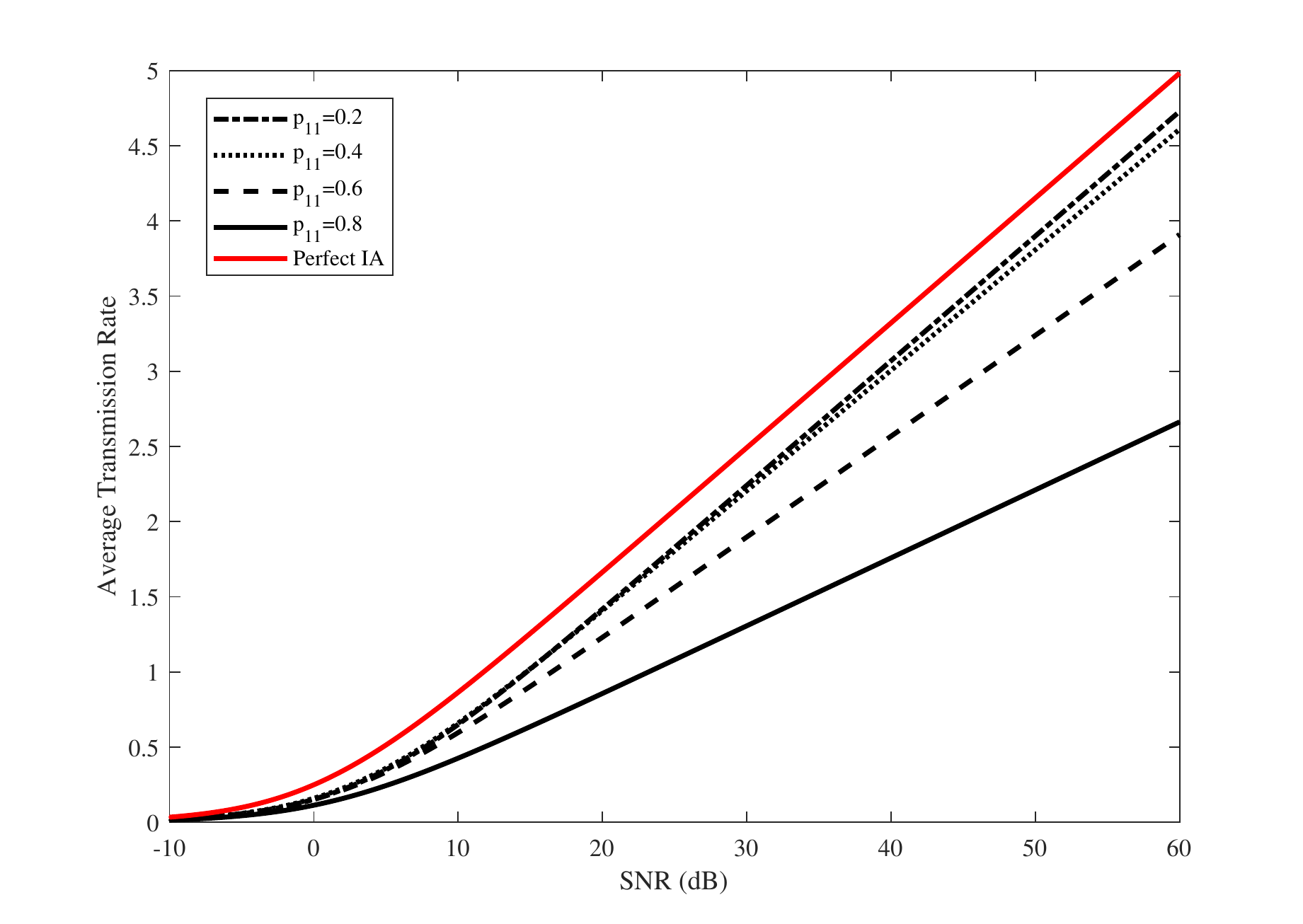}
  \caption{Average achievable rate for each user in the $K-$user IC for $p_{ii}=\{0.2,0.4,0.6,0.8\}$, constant-value cross links, and $n=20$. The red solid line indicates the achievable rates for IA with perfect CSI.}
  \label{TransmissionRate}
\end{figure}


In this subsection, to demonstrate the gain of our proposed strategy, we consider a relatively short transmission block-length, $n=20$, and numerically evaluate the results. 
Fig. \ref{TransmissionRate} shows the average achievable rates for different values of $p_{ii}$ using our strategy. This figure indicates that when $p_{ii}$ decreases and direct channels have a higher variation rate, the chance of finding proper channel states with more interference-free dimensions increases. In other words, when the direct channels have higher variation probability, we can achieve a higher average transmission rate. This figure also includes the achievable rates of IA with perfect CSI. Although, IA with perfect CSI is meant for asymptotic analysis, \emph{i.e.} degrees-of-freedom, as discussed in Section~\ref{Section:Decoding}, we eliminated the channel gains from our analysis, therefore, allowing a comparison to this perfect CSI case.


\section{Conclusion}
\label{Section:Conclusion_IFB}
We proposed a new blind interference alignment strategy for the $K$-user ICs that exploits the radiation pattern fluctuations of transmit antennas. Our strategy includes a multi-layer encoding strategy at the transmitters, an interference alignment phase and a successive decoding phase at each receiver. Our strategy does not rely on any channel state information beyond the CDF of channel variations. We evaluated the achievable rates numerically and compared it to the performance of IA techniques with perfect instantaneous CSI at the transmitters. 
We divided the total transmission power among many layers to achieve the highest average achievable sum rate. 
The next steps include obtaining a closed-form expression for the achievable rates, and extending the results to the scenario in which cross links vary at similar rates when compared to direct links.

\bibliographystyle{ieeetr}
\bibliography{bib_BIA}


\end{document}